\documentstyle[twoside]{article}

\begin{document}
\

\begin{center}
Los Alamos Electronic Archives: quant-ph/9809056 v3\\
\medskip
{\scriptsize  v2 in 
``Symmetries in Quantum Mechanics and Quantum Optics"\\ Burgos, 
Spain, Sept. 21-24, 1998\\
Proceedings edited by F.J. Herranz, A. Ballesteros, 
L.M. Nieto, J. Negro, C.M. Pere\~na\\ 
Servicio de Publicaciones de la Universidad de Burgos, Burgos, Spain,
1999, pp. 301-315\\
ISBN 84-95211-08-4}
\end{center}

\begin{center}

{\bf Short Survey of Darboux Transformations 
}
\medskip


\vskip1cm

{\rm Haret C. Rosu\\
rosu@ifug3.ugto.mx}
\vskip0.5cm

{\it Institute of Physics, Univ. of Guanajuato,
Apdo Postal E-143, Le\'on, Gto, Mexico}

\vskip0.3cm

\end{center}
\bigskip

\begin{abstract}
\noindent
A selective chronological survey of Darboux transformations as related to
supersymmetric
quantum mechanics, intertwining operators and inverse scattering
techniques is presented. Short comments are
appended to each quotation and basic concepts are explained in order to
provide a useful primer.
\end{abstract}


\bigskip

\noindent
{\bf Contents}

\noindent
Chronological set 1: DTs (as covariance of Sturm-Liouville problems)

\noindent
Chronological set 2: Intertwining (transformation) operators (the more
general operator language)

\noindent
Chronological set 3: Related inverse quantum scattering approaches

\noindent
{\bf Some abbreviations}:

\noindent
1. Darboux transformation(s) - DT(s) \hfill
2. Factorization method - FM

\noindent
3. Inverse Quantum Scattering - IQS \quad \quad \quad $\;\;$ $\;$
4. One-dimensional - 1D

\noindent
5. Quantum mechanics - QM \hfill 6. Schroedinger equation - SE 

\noindent
7. Supersymmetry, supersymmetric - SUSY\\

\noindent
Possible combinations are allowed.







\bigskip

\newpage

\begin{center} {\bf CHRONOLOGICAL SET 1

Darboux Transformations

(covariance of Sturm-Liouville problems)} \end{center}


\noindent
{\bf 1882}:
{\em  Sur une proposition relative aux \'equations lin\'eaires}\\

\noindent
This proposition, which history proved to be a
notable {\em theorem}, is provided by G. Darboux in
Compt. Rend. Acad. Sci. {\bf 94}, 1456 (physics/9908003),
and is also to be found in  his course ``Th\'eorie des Surfaces", II, 
p. 210 (Gauthier-Villars, 1889).
Darboux's result was not used and recognized as important for a long time,
and in fact has been {\em only} mentioned as an exercise in 1926 by Ince.\\

\noindent
{\bf 1926}: {\em Ordinary Differential Equations} (Dover)\\

\noindent
In his famous Dover book, E.L. Ince publishes Darboux's theorem in a section of
``Miscellaneous Examples" at page 132, together with other two particular
examples of the theorem related to the free particle and P\"oschl-Teller
potential (the latter belonging to Ince), Exercises 5, 6 and 7, respectively. 
The exercises in Ince follow closely the formulation of Darboux (see \S 1).


\S 1: \underline{Exercise 5 at page 132 of Ince's textbook}\\
Prove that if the general solution $u=u(x)$ of the equation
$$
\frac{d^2u}{dx^2}=(\phi (x)+h)u
$$
is known for all values of $h$, and that any particular solution for the
particular value $h=h_{1}$ is $u=f(x)$, then the general solution of the
equation
$$
\frac{d^2y}{dx^2}=\left(f(x)\frac{d^2}{dx^2}(f^{-1}(x))+h-h_{1}\right)y
$$
for $h\neq h_{1}$ is
$$
y=u^{'}(x)-u(x)\frac{f^{'}(x)}{f(x)}
$$
\\

\noindent
{\bf 1940-1941}:  {\em The factorization of the hypergeometric equation}\\

\noindent
In Proc. Roy. Irish Acad. {\bf 47} A, 53 (1941) (physics/9910003), 
E. Schr\"odinger
factorizes the hypergeometric equation, finding that there are several ways
of factorizing it. This was a byproduct of his FM (published during 1940-1941
in the same review) originating ``from a, virtually, well-known treatment of
the oscillator", i.e., an approach that can be traced back to Dirac's creation
and annihilation operators for the harmonic oscillator and to older
factorization ideas in the mathematical literature.\\

\noindent
{\bf 1951}:  {\em The FM}\\

\noindent
L. Infeld and T.E. Hull present the classification of their
factorizations of linear second order differential equations in
Rev. Mod. Phys. {\bf 23}, 21.\\

\noindent
{\bf 1955}:  {\em Associated Sturm-Liouville systems}\\

\noindent
M.M. Crum publishes an important iterative generalization of Darboux's result
in Quart. J. Math. {\bf 6}, 121 (physics/9908019), 
without any mention of Darboux
(see \S 2).\\


\S 2: \underline{Crum's iteration}\\

\noindent
Let $\psi _1, \psi _2,...\psi _{N}$ be solutions of a given Schr\"odinger
equation $-D^2\psi _{i}+u\psi _{i}=\lambda \psi _{i}$,
for fixed, {\em arbitrary} constants $\lambda=\lambda _{1},
\lambda _2 ,...\lambda _{N}$, respectively.
Define the Wronskian determinant $W$ of $k$ functions $f_1,f_2,..., f_3$ by
$$
W(f_1,..., f_2)=det A,\qquad A_{ij}=\frac{d^{i-1}f_{j}}{dx^{i-1}},\qquad
i,j=1,2,...,k~.
$$
Then, the transformations (Crum's formulas)
$$
\psi[N]=W(\psi _{1},...,\psi _{N}, \psi)/W(\psi _{1},..., \psi _{N})
$$
$$
u[N]=u-2D^{2}\ln W(\psi _{1},..., \psi _{N})
$$
are covariant ones, i.e., one can write the following SL equation:
$$
-D^{2}\psi[N]+u[N]\psi[N]=\lambda\psi[N]~.
$$
Darboux's result of 1882 may be seen as the case $N=1$. 
In other words, Crum presented
the succesive (iterative) DTs in compact formulas.\\

\noindent
{\bf 1957}: {\em Krein's approach to generalized DT}\\

\noindent
M.G. Krein publishes extensions of Crum's iteration in
D.A.N. SSSR {\bf 113}, 970, using a different perspective and with
more elaborations. In particular he provides an important theorem on the
sign of the Wronskian.\\

\noindent
{\bf 1979}: {\em Concept of DT}\\

\noindent
V.B. Matveev uses for the first time the expression DT
(directly in the title) in his series of papers in Lett. Math. Phys. {\bf 3},
217, 425, in which he extended the range of applicability of the concept
to differential-difference and difference-difference evolution equations.
\\

\noindent
{\bf 1981}: {\em Dynamical breaking of SUSY}\\

\noindent
In section 6, ``Some models", of his renowned paper Nucl. Phys. B {\bf 185},
513, E. Witten introduces SUSY QM as a toy model for supersymmetry breaking
in quantum field theories. As a matter of fact, the whole SUSY breaking
paradigm is nothing but a classification of vacuum states based on
the SUSY QM toy model. The SUSY breaking is considered as a sort of
``phase transition" with the order parameter being the Witten index, defined
as the grading operator $\tau=(-1)^{\hat N _{f}}$, where $\hat N _{f}$ is the
fermion number operator.
For the 1D SUSY QM Witten's index is the third Pauli matrix $\sigma _{3}$,
which is +1 for the bosonic sector and -1 for the fermionic sector of the
1D quantum problem at hand.
It is also quite common to call a particular Riccati solution
as a (Witten) ``superpotential". Papers that now are standard references
are published during 1982-1984.
It is known that H. Nicolai was the
first to write the simple SUSY QM matrix algebra in 1976, whereas
a relativistic graded algebra has been first written by Yu.A. Gol'fand
and E.P. Likhtman in JETP Lett. {\bf 13}, 323 (1971) and since then
supersymmetric gauge theories have been main stream research. \\

\noindent
{\bf 1981}: {\em DTs and nonlinear evolution equations}\\

\noindent
V.B. Matveev and M. Salle apply for the first time DT to nonlinear equations
in D.A.N. SSSR {\bf 261}, 533. At present, there are hundreds of papers in
the area.

\S 3: \underline{DTs and B\"acklund Transformations}\\

\noindent
A well-know transformation in nonlinear physics is the B\"acklund
transformation (BT)
[A.V. B\"acklund, Math. Ann. {\bf 9} (1876), 297; {\bf 19} (1882), 387],
that has been first applied to sine-Gordon equation.
When BT operators are applied to N soliton solutions one gets
N+1 soliton solutions. In the Korteweg-de Vries case, DTs may be seen
as BTs for instantaneous solitons, since instantaneous KdV ${\rm sech}^{2}$
solitons may be considered at the same time as Schr\"odinger potentials
of a well-known SUSY QM problem.

\noindent
On the other hand, in the Lax representation of KdV equation, the first Lax 
operator
is of Schroedinger type, a fact showing the importance of Darboux covariance
in nonlinear physics.\\ 

\noindent
{\bf 1983}:
{\em Derivation of exact spectra of the SE by means of
SUSY}\\

\noindent
In JETP Lett. {\bf 38}, 356,
L.E. Gendenshtein introduces the important concept
of {\em shape invariance} (SI) in SUSY QM.

\S 4: \underline{The SI property}\\

\noindent
SI is a property of some classes of
potentials with respect to their parameter(s), say $a$, and reads
$$
V_{n+1}(x,a_{n})=V_{n}(x,a_{n+1})+R(a_{n})
$$
where $R$ is a remainder. This property assures a fully algebraic scheme
for the spectrum and wavefunctions. Fixing $E_{0}=0$, the excited spectrum is
given by
$$
E_{n}=\sum _{k=2}^{n+1} R(a_{k})
$$
and the wavefunctions are obtained from
$$
\psi _{n}(x, a_1)=\prod _{k=1}^{n} A^{+}(x, a_{k})  \psi _{0}(x, a_{n+1})
$$
\\

\noindent
{\bf 1984}:  {\em FM and new potentials with the
oscillator spectrum}\\

\noindent
In J. Math. Phys. {\bf 25}, 3387,
B. Mielnik provides the first application of the general Riccati solution
to the harmonic oscillator, noticing also the similarity to
the Abraham-Moses class of isospectral potentials in the area of
inverse scattering. In the same year, D. Fern\'andez gives a second
application to the atomic hydrogen spectrum in Lett. Math. Phys. {\bf 8}, 337,
whereas M.M. Nieto
clarifies further the inverse scattering aspect of Mielnik's construction
in Phys. Lett. B {\bf 145}, 208.
The procedure may be seen as a double Darboux transformation in
which the general Riccati (superpotential) solution is involved (see \S 7).
Thereby it will be denoted by the acronym DDGR
(double Darboux general Riccati).
\\

\noindent
{\bf 1984-1985}:
{\em FM and DTs for multidimensional Hamiltonians}\\

\noindent
In papers published in Teor. Mat. Fiz. {\bf 61} and subsequent Phys. Lett. A 
versions, A.A. Andrianov, N.V. Borisov and M.V. Ioffe
discover the relation
between SUSY QM and DTs while playing with matrix Hamiltonians in
SUSY QM. \\

\noindent
{\bf 1985}:
{\em Exactness of semiclassical bound state energies for SUSY QM}\\

\noindent
In Phys. Lett. B {\bf 150}, 159,
A. Comtet, A. Bandrauk and D.K. Campbell were
the first to study the WKB features
of SUSY QM introducing a SUSY QM WKB
formula $\int _{a}^{b}[E-W^2(y)]^{1/2}dy=n\pi \hbar$, where $W$ is Witten's
superpotential and $a$ and $b$ are turning points. This research line
has a rich publishing output.\\

\noindent
{\bf 1988}:
{\em PARASUSY QM}\\

\noindent
V.A. Rubakov and V.P. Spiridonov introduce
PARASUSY QM in Mod. Phys. Lett. A {\bf 3}, 1337, involving supercharges
of order-three nilpotency. Later, J. Beckers and N. Debergh found a
different algebra of the same type [Nucl. Phys. B {\bf 340}, 767 (1990)].
There are many generalizations, making this ten-year topic quite active
and interesting from the application point of view. \\

\noindent
{\bf 1991}:  {\em DTs and Solitons} (Springer)\\

\noindent
V.B. Matveev and M. Salle publish the first (excellent) book
(112 pp, 197 refs.)
focusing on DTs and their relation with soliton (mathematical) physics.
The DTs are defined as covariant properties of the SE. Many types of DTs
(not mentioned here) are presented in this book in a concise manner.\\

\S 5: \underline{ Darboux Covariance}\\

\noindent
The statement of the Darboux theorem can be interpreted as the
{\em Darboux covariance} of a Sturm-Liouville equation
$$
-\psi _{xx}+u\psi=\lambda \psi
$$
by which one should understand that the following DT
$$
\psi \rightarrow \psi [1]=(D-\sigma _{1})\psi=\psi _{x}-\sigma _{1}\psi=
$$
$$
\frac{W(\psi _{1},\psi)}{\psi _{1}}
$$
$$
u\rightarrow u[1]=u-2\sigma _{1x}=u-2D^{2}\ln \psi _{1}
$$
where $\sigma _{1}=\psi _{1x}\psi ^{-1}$, i.e. it is the logarithmic
derivative, and $W$ is the Wronskian determinant
passes the SL equation to the (Darboux isospectral) form
$$
-\psi _{xx}[1] +u[1]\psi[1]=\lambda\psi[1]
$$
When DTs are applied iteratively one gets Crum's result.
One can also say that the two SL equations are related by a DT.\\

\noindent
{\bf 1994}: {\em A modification of Crum's method} \\

\noindent In Theor. Math. Phys. {\bf 101}, 1381,
V.E. Adler discusses a double Darboux transform for consecutive
eigenfunctions $\psi _{k}$ and $\psi _{k+1}$ as transformation functions,
producing a {\em nonsingular} potential missing the two
levels $E=k$ and $E=k+1$ in its discrete spectrum (thus anharmonic).
According to Samsonov this result
is implicit in Krein's theorem on the sign of the Wronskian.
Very recently, D.J. Fern\'andez, V. Hussin and B. Mielnik
[Phys. Lett. A {\bf 244}, 309, (1998)] gave an interesting combination of 
Adler's method and DDGR.\\

\noindent
{\bf 1994}: {\em Coherent states for isospectral oscillator Hamiltonians}\\

\noindent In J. Phys. A {\bf 27}, 3547,
D.J. Fern\'andez, V. Hussin and L.M. Nieto provide the first
discussion of coherent states for Darboux transformed systems.
In this important topic there are recent significant results due to
Bagrov and Samsonov [J. Phys. A {\bf 29}, 1011 (1996)] and Samsonov
[J. Math. Phys. {\bf 39}, 967 (1998)].
In the latter paper, Samsonov showed that in some cases the distorsion
of the phase space due to DT is calculable.\\

\noindent
{\bf 1995}: {\em SUSY and QM}\\

\noindent
In Phys. Rep. {\bf 251}, 267,
F. Cooper, A. Khare and U. Sukhatme publish the latest (at this time) review
on SUSY QM. Even though this is a comprehensive work with 265 references,
they cite 16 omitted topics and are forced to accept the following:
``So much work has been done in the area of SUSY QM in the last 12 years
that it is almost impossible to cover all the topics in such a review".
For comparison, what may be considered (in a limited sense)
as the first SUSY QM review paper
written by L.E. Gendenshtein and I.V. Krive in 1985 has 65 references in
various areas.\\

\noindent
{\bf 1996}: {\em SUSY Methods in Quantum and Statistical
Physics} (Springer)\\

\noindent
G. Junker publishes his {\em Habilitazion} at Erlangen as a book (124 pp)
with more than 300 references.\\

\noindent
{\bf 1996}: {\em DTs for time-dependent SE (TDSE)}\\

\noindent
In Phys. Lett. A {\bf 210}, 60,
V.G. Bagrov and B.F. Samsonov provide an important extension of DTs to
nonstationary SEs by means of intertwining.
A direct (less general) factorization approach for the time-dependent
Pauli equation has been given by V.A. Kosteleck\'y, V.I. Man'ko, M.M. Nieto,
and D. Truax, in Phys. Rev. A {\bf 48}, 951 (1993);
see also, V.M. Tkachuk, J. Phys. A {\bf 31}, 1859 (1998).\\

\noindent
{\bf 1997}: {\em DT for Dirac eqs with (1+1) potentials}\\

\noindent
A.V. Yurov studies DT for Dirac equations in Phys. Lett. A {\bf 225}, 51.\\

\noindent
{\bf 1997}: {\em DT of the SE}\\

\noindent
In Phys. Part. Nucl. {\bf 28}, 374 (1997)
[Fiz. Elem. Chastits At. Yadra {\bf 28}, 951 (1997)]
V.G. Bagrov and B.F. Samsonov publish a first review of the results in QM
obtained by means of DT. There are 83 references.\\


\S 6: \underline{DTs and SUSY QM} (Matveev and Salle)\\

\noindent
{\bf Proposition}: {\em Witten's SUSY QM is equivalent to a single DT.}\\

\noindent
{\bf Proof}: Consider two Schr\"odinger equations
$$
-D^2 \psi +u\psi=\lambda \psi
$$
$$
-D^2 \phi +v\phi=\lambda \phi
$$
related by DT, i.e., $v=u[1]$ and $\phi =\psi [1]$.

\noindent
Notice now that the function $\phi _{1}=\psi ^{-1}_{1}$ satisfies the
Darboux-transformed equation for $\lambda =\lambda _{1}$.

\noindent
If now one uses the second (transformed) equation as initial one and
perform the DT with the generating function $\phi _{1}$, one just go
back to the initial $u$ equation.
Thus, one can define an {\em inverse DT}, that follows
in a clear way from the direct one:
$$
u=v-2D^{2}\ln \phi _{1}=v[-1]=v-2D^{2}\ln \psi ^{-1}_{1}
$$
$$
\psi =\left(\phi _{x}-\frac{\phi _{1x}}{\phi _{1}}\phi\right)
(\lambda _{1}-\lambda)
=\left(\phi _{x}+\frac{\psi _{1x}}{\psi _{1}}\phi\right)
(\lambda _{1}-\lambda)
$$
If the {\em sigma} notation for the logarithmic derivative is introduced,
i.e.,
$$
\sigma =\frac{\psi _{1x}}{\psi _{1}}=-\frac{\phi _{1x}}{\phi _{1}}
$$
the Riccati (SUSY QM) representation of the Darboux pair of Schr\"odinger
potentials is obtained
$$
u=v[-1]=\sigma _{x}+\sigma ^{2}+\lambda _{1}
$$
$$
v=u[1]=-\sigma _{x}+\sigma ^{2}+\lambda _{1}
$$
It is now easy to enter the SUSY QM concept of supercharge operators.
For that, one employs the factorization operators
$$
B^{+}=-D+\sigma,\qquad B^{-}=D+\sigma
$$
They effect the wavefunction part of the direct and inverse
DT, respectively.
Moreover
$$
B^{+}B^{-}=-D^{2}+v-\lambda _{1}
$$
$$
B^{-}B^{+}=-D^{2}+u-\lambda _{1}
$$
Thus, the commutator $[B^{+},B^{-}]=v-u=-2D^{2}\ln \psi _{1}$ gives the
Darboux difference in the shape of the Darboux-related
potentials. Introducing the Hamiltonian operators
$$
H^{+}= B^{-}B^{+}+\lambda _{1}
$$
$$
H^{-}=B^{+}B^{-}+\lambda _{1}
$$
one can also interpret the $B$ operators as factorization ones
and write the famous matrix representation of SUSY QM, as well as
the simplest possible superalgebra.

\noindent
The factorizing operators in matrix representation are called
{\em supercharges} in SUSY QM, and are nilpotent operators
$$ Q^{-}= A_-\sigma _+ =
\left( \begin{array}{cc}
0 & 0 \\
A^{-} & 0
\end{array} \right)$$
and
$$Q^{+}= A_+\sigma _-  =
\left( \begin{array}{cc}
0 & A^{+} \\
0 & 0
\end{array} \right)$$

\noindent
$\sigma _-= \left( \begin{array}{cc}
0 & 1 \\
0 & 0
\end{array} \right)$ and
$\sigma _+=\left( \begin{array}{cc}
0 & 0 \\
1 & 0
\end{array} \right)$ are Pauli matrices.
In this realization, the matrix form of the
Hamiltonian operator reads
$$ {\cal H} =
\left( \begin{array}{cc}
A^+ A^- & 0 \\
0 & A^- A^+
\end{array} \right) =
\left( \begin{array}{cc}
H_- & 0 \\
0 & H_+
\end{array} \right)
$$
defining the partner Hamiltonians as diagonal elements of the matrix one.
They are partners in the sense that they are isospectral, apart from
the ground state $\phi _{gr,-}$ of $H_-$, which is not included in the
spectrum of $H_+$.

\noindent
{\bf Remark}:
Denoting $\Lambda _{1}(x)=-2D^2\ln \psi _{1}(x)+\lambda _{1}$, one can see
that for a Gaussian of width $\sigma ^2$ one
gets $\Lambda _{1}=2\sigma ^{-2}+\lambda _{1}$, and in the case of
Gaussians of constant width (harmonic ground states)
$\Lambda _{1}$ is just a constant shift of the factorization constant.\\

\bigskip

\setcounter{equation} {0}
\S 7: \underline{The DDGR Construction} (Rosu)\\
\vskip 2mm

\noindent
DDGR offers  an interesting possibility to construct families of potentials
{\em strictly} isospectral with respect to the initial (bosonic) one, if
one asks for the most general superpotential (i.e., the general
Riccati solution)
such that $\rm V_+(x)=  w_{g}^2 + \frac{d w_{g}}{dx}$, where ${\rm V_+}$ is
the fermionic partner potential. It is easy to see that
one particular solution to this equation is ${\rm w_p= w(x)}$, where w(x) is
the common Witten superpotential. One is led to consider the following
Riccati equation ${\rm  w_{g}^2 + \frac{d w_{g}}{dx}=w^2_p +\frac{d w_p}{dx}}$,
whose general solution can be written down as 
${\rm w_{g}(x)= w_p(x) + \frac{1}{v(x)}}$, where ${\rm v(x)}$ is an unknown
function. Using this ansatz, one obtains for the function ${\rm v(x)}$ the
following Bernoulli equation
$$
{\rm \frac{dv(x)}{dx} - 2 \, v(x)\, w_p(x) = 1},
\eqno(1)
$$
that has the solution
$$
{\rm v(x)= \frac{{\cal I}_0(x)+ \lambda}{u_{0}^{2}(x)}},
\eqno(2)
$$
where ${\rm {\cal I}_0(x)= \int _{c}^{x} \, u_0^2(y)\, dy}$, ($c=-\infty$
for full line problems and $c=0$ for half line problems, respectively),
and $\lambda$ is an
integration constant thereby considered as a free DDGR parameter.
Thus, ${\rm w_{g}(x)}$ can be written as follows
$$
{\rm w_{g}(x;\lambda)=  w_p(x) + \rm \frac{d}{dx}} \Big[ {\rm ln}
({\cal I}_0(x) + \lambda) \Big]
\eqno(3a)
$$
$$
={\rm w_p(x)+\sigma _{0}(\lambda)}
\eqno(3b)
$$
$$
={\rm - \frac{d}{dx} \Big[ ln \left(\frac{u_0(x)}{{\cal I}_0(x) +
\lambda}\right)\Big]}.
\eqno(3c)
$$
Finally, one easily gets the $V_-(x;\lambda)$ family of
potentials 
$$
{\rm  V_-(x;\lambda)} = {\rm w_{g}^2(x;\lambda) -
\frac{d w_{g}(x;\lambda)}{dx}}
\eqno(4a)
$$
$$
= {\rm V_-(x) - 2 \frac{d^2}{dx^2} \Big[ ln({\cal I}_0(x) + \lambda)}
\Big]
\eqno(4b)
$$
$$
= {\rm V_-(x) -2\sigma _{0,x}(\lambda)}
\eqno(4c)
$$
$$
= {\rm V_-(x) - \frac{4 u_0(x) u_0^\prime (x)}{{\cal I}_0(x)
+ \lambda} 
+ \frac{2 u_0^4(x)}{({\cal I}_0(x) + \lambda)^2}.}
\eqno(4d)
$$
All ${\rm  V_-(x;\lambda)}$ have the same supersymmetric partner potential
${\rm V_+(x)}$ obtained by deleting the ground state.
They are asymmetric double-well potentials that may be considered as a sort
of intermediates between the bosonic potential ${\rm V_-(x)}$ and
the fermionic partner ${\rm V_+(x)=V_-(x)-2\sigma _{0,x}(x)}$.
From Eq. ($3c$) one can infer the ground state wave functions
for the potentials ${\rm V_-(x;\lambda)}$ as follows
$$
{\rm u_0(x;\lambda)= f(\lambda)
\frac{u _0(x)}{{\cal I}_0(x) + \lambda}},
\eqno(5)
$$
where ${\rm f(\lambda)}$ is a normalization factor that can be shown to be
of the form
${\rm f(\lambda)= \sqrt{\lambda(\lambda +1)}}$.
One can now understand the double Darboux feature of the DDGR by
writing the parametric family in terms of their unique ``fermionic" partner
$$
{\rm V_{-} (x;\lambda)=V_{+}(x)
-2\frac{d^2}{dx^2}\ln\left(\frac{1}{u_{0}(x;\lambda)}\right)},
\eqno(6)
$$
which shows that the DDGR transformation is of the
inverse Darboux type, allowing at the same time
a two-step (double Darboux) interpretation,
namely, in the first step one goes to the fermionic system and in the
second step one returns to a deformed bosonic system.


\bigskip

\begin{center}{\large {\bf CHRONOLOGICAL SET 2

Intertwining (Transformation) Operators

(the more general operator language)}}\end{center}

\noindent
{\bf 1938}: {\em Intertwining discovered}\\

\noindent
J. Delsarte introduces the notion of transformation (transmutation)
operator in
{\em Sur certaines transformationes fonctionelles relative aux \'equations
lineaires aux derivees partielles du 2nd ordre},
Comp. Rend. Acad. Sci. (Paris) {\bf 206}, 1780 (physics/9909061),
[see also, {\em Sur une extension de la formule de Taylor}, J. Math. Pures
et Appl. {\bf 17}, 213 (1938)], but it is only
18 years later that he with J.L. Lions expound more on the method.
J. Delsarte, Colloq. Int. Nancy, p. 29 (1956); J.L. Lions, ibidem, p. 125
(1956); J. Delsarte, J.L. Lions, Comment. Math. Helv. {\bf 32}, 113 (1956).\\

\S 8 \underline{What is intertwining ?}\\

\setcounter{equation} {0}
\noindent
Two operators $L_{0}$ and $L_{1}$ are said to be intertwined by an operator
$T$ if
\begin{equation} \label{14}
L_{1}T=TL_{0}~.
\end{equation}
If the eigenfunctions $\varphi _{0}$ of $L_0$ are known, then from the
intertwining relation one can show that the (unnormalized) eigenfunctions
of $L_1$ are given by $\varphi _{1}=T\varphi _{0}$.  The main problem in
the intertwining transformations is to construct the transformation
operator $T$. 1D QM is one of the simplest
examples of intertwining relations since Witten's transformation operator
$T_{qm}=T_1$ is just a first spatial derivative
plus a differentiable coordinate function (the superpotential) that should
be a logarithmic derivative of the true bosonic zero mode (if it exists),
but of course
higher-order transformation operators can be constructed without much
difficulty.

\noindent
Thus, within the realm of 1D QM, writing $T_1=
D-\frac{u^{'}}{u}$, where $u$ is a true bosonic zero mode, one can
infer that the
adjoint operator $T^{\dagger}_{1}=-D-\frac{u^{'}}{u}$ intertwines in
the opposite direction, taking solutions of $L_{1}$ to those of $L_{0}$
\begin{equation}  \label{15}
\varphi _{0}=T_{1}^{\dagger}\varphi _{1}~.
\end{equation}
In particular, for standard 1D QM, $L_0=H_{-}$ and $L_1=H_{+}$ and
although the true zero mode of $H_{-}$ is annihilated by $T_1$, the
corresponding (unnormalized) eigenfunction of $H_{+}$ can nevertheless
be obtained by
applying $T_1$ to the other independent zero energy solution of $H_{-}$.\\

\noindent
{\bf 1973}: {\em Theory of generalized shift operators} (Nauka)\\

\noindent
This is one of the remarkable books of B.M. Levitan edited by the publishing
house Nauka.\\

\noindent
{\bf 1978}: {\em Applications of a commutation formula}\\

\noindent
P.A. Deift presents applications of the so-called ``commutation formula"
(that can be found, e.g., in the book of S. Sakai,
{\em $C^{*}$-Algebras and $W^{*}$-Algebras}, Springer, 1971). In the
paper of Deift in Duke Math. J. {\bf 45}, 267 (1978), section 4 contains
the application of the commutation formula to ordinary differential operators.
Many of the results in that section have been reproduced later in
SUSY QM style, e.g. by Sukumar.\\

\noindent
{\bf 1986-1987}:  {\em Isometric operators, isospectral Hamiltonians, and
SUSY QM}\\

\noindent
In Phys. Rev. D {\bf 33}, 2267, D {\bf 36}, 1103,
D.L. Pursey uses intertwining in his combined procedures of generating
families of strictly isospectral Hamiltonians, starting from the Marchenko
inverse scattering equation.\\

\noindent
{\bf 1991}: {\em Intertwining of exactly solvable Dirac eqs. with 1D
potentials}\\

\noindent
In Phys. Rev. A {\bf 43}, 4602,
A. Anderson applies matrix intertwining relations to the Dirac equation
showing that their structure is described by an N=4 superalgebra.\\



\noindent
{\bf 1995-1998}: {\em Intertwining widely used}\\

\noindent
Intertwining is already well known to many active authors in SUSY QM,
who are playing with higher-order generalizations. But, as
always, the most important (at least for standard quantum mechanics) are
the simplest cases, namely the Darboux first-order intertwining operators.\\

\setcounter{equation} {0}
\S 9:  \underline{Using intertwining: DTs for TDSE} (Bagrov and Samsonov)\\

\noindent
The DT for the TDSE are based on
the intertwining relation
$$
T(i\partial _{t}-H_0)=(i\partial _{t}-H_1)T\,,
\eqno(1)
$$
where
$$
H_i=-\partial _{x}^2+V_i(x,t)\,,\quad i=0,1\,,
\eqno(2)
$$
and $T$ is a first-order diff. transformation operator of the form
$$
T=L_1(x,t)\partial _{x}+L_0(x,t)\,.
\eqno(3)
$$
It follows immediately from the intertwining
relation that if $\psi _0$ solves the TDSE with Hamiltonian
$H_0$, then $\psi_1=T\psi_0$ will solve the TDSE with Hamiltonian $H_1$.
It is also easily verified that the intertwining relation
will be satisfied if and only if 
$$
T=L_{1}(\partial _{x}+\chi_{x})\,,\qquad
V_{1}=V_{0}+2\chi_{xx}+i(\log L_{1})_{t}\,,
\eqno(4)
$$
where $e^{-\chi}$ is a solution of the TDSE with potential $V_0$,
and $L_1=L_1(t)$ is an arbitrary function. The transformed potential
$V_1(x,t)$ is a real-valued function if and only if
$$
{\rm Im}\chi_{xxx}=0
$$
and
$$
|L_1|=\exp\left[-2\int^t_{t_0} {\rm Im}\chi_{xx}(x,s)\,ds\right]\,.
$$
Without loss of generality, one can assume that $L_1$
is real and positive, and is therefore given by the right-hand side
of the above equation.

\noindent
Just as in the time-independent case, the DT
for the TDSE can be inverted. Thus, if $\psi_1$ is a solution of
the TDSE with potential $V_{1}$ given by Eq.(4), the function
$$
\psi_0(x,t)=\frac{e^{-\chi(x,t)}}{L_1(t)}\left[\int_{x_0}^x
e^{\chi(y,t)}\psi_1(y,t)dy+ c_0(t)\right]
\eqno(5)
$$
with $c_0(t)$ given by\\
$$
c_0(t)=iL_1(t)\int_{t_0}^t \frac{e^{\chi(x_0,s)}}{L_1(s)}
\big(\psi_{1,x}(x_0,s)-\chi_x(x_0,s)\psi_1(x_0,s)
\big)\,ds
$$
is a solution of the TDSE with potential $V_0$.
As remarked by F. Finkel {\em et al.} in math-ph/9809013 if the
factor $L_1$ is taken
as unity, the mapping $\psi_1\mapsto\psi_0$ given by Eq.(5)
reduces to the non-local transformation
considered by Bluman and Shtelen in  J. Phys. A {\bf 29}, 4473 (1996).\\


\begin{center}{\large {\bf CHRONOLOGICAL SET 3

Related Inverse Quantum Scattering (IQS) Approaches}}\end{center}

\noindent
In the case of classical dynamics, the inverse problem just means to
determine the force acting on a macroscopic body from the features of the
trajectory. In this sense, one of the oldest inverse problem has been
Newton's problem of determining the force on the planets from the Kepler
properties (laws) of their movement.

\noindent
In the realm of differential operators,
an inverse spectral method generally  means to determine the
(linear) operator from some given spectral data. Roughly speaking, in the
case of a Sturm-Liouville operator, this would reduce to getting the
potential. Thus, there is a certain parallel to the classical case.
As well known, in the IQS approach one uses integral transformation operators.
The key point for the construction of the integral operators are the famous
Gel'fand-Levitan equation and Marchenko equation. The IQS research started in
the second half of the 1940s and the 1950s have been a real boom.\\

\noindent
{\bf 1954-1956}: {\em Krein's approach to IQS}\\

\noindent
M.G. Krein reports a new IQS approach in three D.A.N. notes,
DAN SSSR {\bf 97}, 21 (1954); {\bf 105}, 433 (1955); {\bf 111}, 1167 (1956).\\

\S 10: \underline{Krein's IQS}\\

\noindent
As explained to us by Chadan and Sabatier in their book (1977, 1989), 
Krein's
approach is the following way to solve the $l=0$ inverse scattering problem.
The SE
$$
-D^2y+V(r)y=k^{2}y~,
$$
where $D=\frac{d}{dr}$, is substituted by the equivalent system
$$
Dy+A(r)y=kz
$$
$$
-Dy+A(r)z=ky~.
$$
The two functions $V$ and $A$ are connected by the Riccati equation
$$
V=-DA+A^{2}~.
$$
From the analytical properties of the Schr\"odinger
regular solution $\phi(k,r)$ and the Wiener-Paley theorem, one can
obtain the following representation
$$
\phi(k,r)=k^{-1}{\rm Im}\Big[e^{ikr}\left(1+\int _{0}^{2r}\Gamma _{2r}(t)
e^{-ikt}dt\right)\Big]~,
$$
where $A(r)=2\Gamma _{2r}(2r)$.
Moreover, for any fixed value of $r$, the function $\Gamma _{2r}(t)$
is a solution of the Fredholm integral equation
$$
\Gamma _{2r}(t)+H(t)+\int _{0}^{2r}\Gamma _{2r}(s)H(s-t)ds=0
$$
where, $H(t)$ is related to the Jost function $F(k)$ in the following way
$$
H(t)=\frac{1}{\pi}\int _{0}^{\infty}\Big[|F(k)|^{-2}-1\Big]\cos kt dk
$$
The scheme to solve the inverse scattering problem according to Krein is
first to build the function $H(t)$, then to get $\Gamma _{2r}(t)$,
leading immediately to $A(r)$, and finally obtaining the potential $V(r)$
from the Riccati equation.\\

\noindent
{\bf 1980}: {\em Changes in potentials due to changes in the point spectrum:
Anharmonic oscillators with exact solutions}\\

\noindent
Using the Gel'fand-Levitan equation,
P.B. Abraham and H.E. Moses introduce a procedure of deleting and adding
bound states, which is ``almost" equivalent to DDGR 
[Phys. Rev. A {\bf 22}, 1333].
\\

\S 11: \underline{Comparison between strictly isospectral techniques within
IQS}\\

\noindent
According to A. Khare and U. Sukhatme [Phys. Rev. A {\bf 40}, 6185 (1989)],
the difference between DDGR, Abraham-Moses, and Pursey schemes when used
in the IQS context to generate one-parameter strictly isospectral potentials
on the base of an initial zero mode $\psi _{0}$
lies merely in the employed function used for the (inverse)
steps of readding the zero-mode as follows
($I=\int _{-\infty}^{x} \psi _{0}^{2}(y)dy$):

\noindent
For DDGR, the function is $\phi=\psi _{0}^{-1}$, being a node-free solution
of the first step transformed equation.

\noindent
For Pursey's method, it is $v=\psi _{0}/I$, such that $v$ is a solution of the
first step zero-energy SE that vanishes at $+\infty$.

\noindent
For the Moses-Abraham procedure, one should use $u=\psi _{0}/(1-I)$, where
$u$ is a solution of the first step zero-energy SE vanishing at $-\infty$.

\noindent
One can obtain five independent families of strictly isospectral
potentials when combining the three procedures.
However, only the DDGR method leads to reflection and
transmission amplitudes identical to those of the original potential, showing
the complete degeneracy produced by such a construction.\\

\noindent
{\bf 1985}: {\em First issue of Inverse Problems}\\

\noindent
Birth year of the IOP review {\em Inverse Problems}.\\

\noindent
There are many good books in the IQS field, e.g., those of
K. Chadan and P.C. Sabatier (1977, 1989), B.M. Levitan (1984),
V.A. Marchenko (1986), 
B.N. Zakhariev and A.A. Suzko (1985).
According to
Levitan, the generalized use of intertwining operators in IQS is due to
the works of V.A. Marchenko.\\

\noindent
{\bf 1985}: {\em SUSY QM and the Inverse Scattering Method}\\

\noindent
In J. Phys. A {\bf 18}, 2937, C.V. Sukumar provides an important
contribution to the understanding of the connections between the two
approaches.\\

\noindent
{\bf 1993}: {\em Bound states in the continuum (BSICs) from SUSY QM}\\

\noindent
In Phys. Rev. A {\bf 48}, 3525, J. Pappademos, U. Sukhatme, A. Pagnamenta 
are the first to apply DDGR to get BSICs. They also study two-parameter
families of SUSY BSIC potentials.\\

\noindent
{\bf 1995}: {\em SUSY transformations of real potentials on the line}\\

\noindent
In J. Phys. A {\bf 28}, 5079,
J.-M. Sparenberg and D. Baye present an exhaustive work on SUSY iterations
showing the power of the method in generating isospectral potentials.\\

\noindent
{\bf 1994}:
{\em Exactly solvable models for the SE from generalized DTs}\\

\noindent
In J. Phys. A {\bf 27}, 2605,
W.A. Schnizer and H. Leeb show in the IQS context that integral
transformations with
a {\em degenerate kernel} are equivalent to differential ones.\\

\noindent
{\bf 1995}:
{\em On the equivalence of the integral and the differential
exact solution generation methods for the 1D SE}\\

\noindent
In J. Phys. A {\bf 28}, 6989, B. Samsonov shows the same
equivalence in a very transparent way, giving an integral form of
Crum's iteration.

{\section*{Conclusion}}
\noindent
A scientific review covering 116 years, even when the chronological style
is used,
cannot have any claim of completeness and people reading it may see
a lot of missing material in general due to the natural biases acting on
the author(s). In my case, a strong bias has been a dead line forcing me to
leave out many active authors and their results in a number of important
topics not quoted above, such as the connections between partially
algebraic quantum problems (quasi-exactly solvable ones) and SUSY QM,
or intricate connections going directly to the core of more sophisticated  
theories.
Nevertheless, even from this brief chronological collection one can get a
partial grasp of what has been done and a feeling of what could be done in
the future.


{\section*{Acknowledgment}}
\noindent 
This work has been partially supported by CONACyT (Mexico) Project
No. 458100-5-25844E. The author wishes to thank L.J. Boya and B. Mielnik
for remarks.

\bigskip







\end{document}